\documentclass[12pt,preprint]{aastex}

\slugcomment{Submitted to ApJ} 

\shorttitle{Rotation Measure due to IGMF}
\shortauthors{Akahori and Ryu}

\begin{document}

\title{Faraday Rotation Measure due to the Intergalactic Magnetic Field}
\author{Takuya Akahori$^1$ and Dongsu Ryu$^2$\altaffilmark{,3}}
\affil{$^1$Research Institute of Basic Science, Chungnam National University,
Daejeon, Korea: akataku@canopus.cnu.ac.kr\\
$^2$Department of Astronomy and Space Science, Chungnam National
University, Daejeon, Korea: ryu@canopus.cnu.ac.kr}
\altaffiltext{3}{Author to whom any correspondence should be addressed.}

\begin{abstract}

Studying the nature and origin of the intergalactic magnetic field (IGMF)
is an outstanding problem of cosmology.
Measuring Faraday rotation would be a promising method to explore the
IGMF in the large-scale structure (LSS) of the universe.
We investigated the Faraday rotation measure (RM) due to the IGMF in
filaments of galaxies using simulations for cosmological structure formation. 
We employed a model IGMF based on turbulence dynamo in the LSS of the universe;
it has an average strength of $\langle B \rangle \sim 10$ nG and a coherence
length of several $\times\ 100\ h^{-1}$ kpc in filaments.
With the coherence length smaller than path length, the inducement
of RM would be a random walk process, and we found that the resultant RM
is dominantly contributed by the density peak along line of sight.
The rms of RM through filaments at the present universe was
predicted to be $\sim 1\ {\rm rad\ m^{-2}}$.
In addition, we predicted that the probability distribution function of
$|{\rm RM}|$ through filaments follows the log-normal distribution,
and the power spectrum of RM in the local universe peaks at a scale
of $\sim 1\ h^{-1}$ Mpc.
Our prediction of RM could be tested with future instruments.

\end{abstract}


\section{Introduction}

The intergalactic medium (IGM) contains gas, which was heated mostly
by cosmological shocks \citep{rkhj03}, along with dark matter;
the hot gas with $T > 10^7$ K is found inside and around clusters/groups
of galaxies and the warm-hot intergalactic medium (WHIM) with
$T = 10^5 - 10^7$ K resides mostly in filaments of galaxies, while
lower temperature gas is distributed mostly as sheetlike structures
or in voids \citep{co99,krcs05}.
As the gas in the interstellar medium, the gas in the intracluster medium
(ICM) and filaments is expected to be permeated with magnetic fields.
Measuring Faraday rotation, the rotation of the plane of linearly-polarized
light due to the birefringence of magneto-ionic medium, has been one of a
few methods to explore the intergalactic magnetic field (IGMF).

Observational exploration of the IGMF using Faraday rotation measure (RM)
was started with the investigation of the intracluster magnetic field
(ICMF) \citep[see][for a review]{car02}.
An RM study of the Coma cluster, for instance, revealed the ICMF of
the strength of order $\sim\mu$G for the coherent length of order
$\sim 10$\ kpc \citep{kim90}.
For Abell clusters, the RM of typically $\sim100 -200\ {\rm rad\ m^{-2}}$
was observed, indicating an average strength of the ICMF to be
$\sim 5$--10\ $\mu$G \citep{cla01,cla04}.
RM maps of clusters were analyzed to study the power spectrum of turbulent
magnetic fields in the ICM; for instance,
a Kolmogorov-like spectrum with a bending at a few kpc scale was found in
the cooled core region of the Hydra cluster \citep{vog05}, and spectra
consistent with the Kolmogorov spectrum were reported in the wider ICM
for the Abell 2382 cluster \citep{gmgp08} and for the Coma cluster
\citep{bfmg10}.

The nature of the IGMF in filaments, on the contrary, remains largely
unknown, because the study of RM outside clusters is still scarce
\citep[e.g.,][]{xkhd06};
detecting the RM due to the IGMF in filaments is difficult with
current facilities, and also removing the galactic foreground is not a
trivial task.
The next generation radio interferometers including the Square Kilometer
Array (SKA), and upcoming SKA pathfinders, the Australian SKA Pathfinder
(ASKAP) and the South African Karoo Array Telescope (MeerKAT), as well the
Low Frequency Array (LOFAR), however, are expected to be used to study
the RM.
Particularly, the SKA could measure RM for $\sim 10^8$ polarized
extragalactic sources across the sky with an average spacing
of $\sim 60$ arcsec between lines of sight (LOS's)
\citep[see, e.g.,][and references therein]{car04,kra09}, enabling us to
investigate the IGMF in the large-scale structure (LSS) of the universe.

Attempts to theoretically predict the RM due to the IGMF have been made:
for instance, \citet{rkb98} and \citet{dol05} used hydrodynamic simulations
for cosmological structure formation to study RM in the LSS, and more
recently \citet{dub08} used MHD simulations to study RM for clusters.
However, the properties of the IGMF, especially in filaments, such as
the strength and coherence length as well as the spatial distribution,
are largely unknown, hindering the theoretical study of RM in the LSS
of the universe.

Recently, \citet{rkcd08} proposed a physically motivated model for the
IGMF, in which a part of the gravitational energy released during
structure formation is transferred to the magnetic field energy as a
result of the turbulent dynamo amplification of weak seed fields
in the LSS of the universe.
In the model, the IGMF follows largely the matter distribution in the
cosmic web and the strength is predicted to be
$\langle B\rangle \sim 10$ nG in filaments.
\citet{cr09} studied various characteristic length scales of magnetic
fields in turbulence with very weak or zero mean magnetic field, and
showed that the coherence length defined for RM is 3/4 times the integral
scale in the incompressible limit.
They predicted that in filaments, the coherence length for RM would be
a few $\times\ 100\ h^{-1}$ kpc with the IGMF of \citet{rkcd08} and the
RM due to the magnetic field would be of order $\sim 1\ {\rm rad\ m^{-2}}$.

In this paper, we study RM in the LSS of the universe, focusing on
RM through filaments, using simulations for cosmological
structure formation along with the model IGMF of \citet{rkcd08} and
\citet{cr09}.
Specifically, we present the spatial distribution, probability
distribution function (PDF) and power spectrum of the RM, and discuss the
prospect of possible observations of the RM.
In sections 2 and 3, we describe our model and the results.
Discussion is in Section 4, and Summary and Conclusion follows in Section 4.

\section{Model}

To investigate RM in the LSS of the universe, we used structure formation
simulations for a concordance $\Lambda$CDM universe with the following
values of cosmological parameters:
$\Omega_{\rm BM}=0.043$, $\Omega_{\rm DM}=0.227$,
$\Omega_{\rm \Lambda}=0.73$,
$h\equiv H_0/(100\ {\rm km\ s^{-1}\ Mpc^{-1}})=0.7$, $n=1$, and
$\sigma_8=0.8$ \cite[same as in][]{rkcd08}.
They were performed using a particle-mesh/Eulerian, cosmological
hydrodynamic code \citep{rokc93}.
A cubic region of comoving volume $(100\ h^{-1}{\rm Mpc})^3$ was
reproduced with $512^3$ uniform grid zones for gas and gravity and
$256^3$ particles for dark matter, so the spatial resolution is
$195\ h^{-1}$ kpc.
Sixteen simulations with different realizations of initial condition were
used to compensate cosmic variance.

For the IGMF, we employed the model of \citet{rkcd08};
it proposes that turbulent-flow motions are induced via the cascade
of the vorticity generated at cosmological shocks during the formation of
the LSS of the universe, and the IGMF is produced as a consequence of the
amplification of weak seed fields of any origin by the turbulence.
Then, the energy density (or the strength) of the IGMF can be estimated
with the eddy turnover number and the turbulent energy density as follow:
\begin{equation}
\varepsilon_B = \phi \left({t \over t_{\rm eddy}}\right)
\varepsilon_{\rm turb}.
\end{equation}
Here, the eddy turnover time is defined as the reciprocal of the vorticity
at driving scales, $t_{\rm eddy} \equiv 1/\omega_{\rm driving}$ 
(${\vec \omega} \equiv {\vec \nabla}\times{\vec v}$), and $\phi$ is the
conversion factor from turbulent to magnetic energy that depends on the
eddy turnover number $t/t_{\rm eddy}$.
The eddy turnover number was estimated as the age of universe times the
magnitude of the local vorticity, that is, $t_{\rm age}\ \omega$.
The local vorticity and turbulent energy density were calculated
from simulations for cosmological structure formation described above.
A functional form for the conversion factor was derived from a separate,
incompressible, magnetohydrodynamic (MHD) simulation of turbulence dynamo.
For the direction of the IGMF, we used that of the passive fields from
simulations for cosmological structure formation, in which weak seed
fields were evolved passively, ignoring the back-reaction, along with
flow motions \citep{kcor97,rkb98}.

In our model, as seed magnetic fields, we took the ones generated through
the Biermann battery mechanism \citep{biermann50} at cosmological shocks.
There are, on the other hand, a number of mechanisms that have been
suggested to create seed fields in the early universe.
Besides various inflationary and string theory mechanisms, the followings
include a partial list of astrophysical mechanisms.
At cosmological shocks, in addition, Weibel instability can operate and
produce magnetic fields \citep{msk06, ss03}, and streaming cosmic rays
accelerated by the shocks can amplify weak upstream magnetic fields
via non-resonant growing mode \citep{bell04}.
In addition, for instance, galactic outflows during the starburst phase
of galactic evolution \citep{ddlm09} and the return current induced by
cosmic-rays produced by Supernovae of first stars \citep{mb10} were
suggested to deposit seed fields.
We point, however, that in our model the IGMF resulting from turbulent
amplification should be insensitive to the origin of seed fields.

The spatial distribution of the strength of the resulting IGMF is shown
in Figure 4 of \citet{rkcd08} and Figure 1 of \citet{rdk10}.
It is very well correlated with the distribution of matter.
The average strength of our model IGMF for the WHIM with
$10^5 < T  <10^7\ {\rm K}$ in filaments is
$\langle B\rangle \sim 10$ nG,
$\langle B^2\rangle^{1/2} \sim$ a few $\times\ 10$ nG,
$\langle \rho B\rangle/\langle \rho\rangle \sim 0.1\ \mu$G,
or $\langle (\rho B)^2\rangle^{1/2}/\langle \rho^2\rangle^{1/2} \sim$
a few $\times~0.1\ \mu$G.

\section{Results}

We calculated RM, defined as $\Delta \chi / \Delta \lambda^2$ ($\chi$ is
the rotation angle of linearly-polarized light at wavelength $\lambda$),
in the local universe with $z=0$ along a path length of $L=100\ h^{-1}$
Mpc, which is the box size of structure formation simulations.
Figure 1 shows the resulting RM map of $(28\ h^{-1}{\rm Mpc})^2$ area in
logarithmic and linear scales.
RM traces the large-scale distribution of matter, and we see two
clusters and a filamentary structure containing several groups.
Through the clusters, groups, and filament in the field, RM is roughly
$\sim 100$, $\sim 10$, and $\sim 1$, respectively, while RM through sheets
and voids is much less.
The bottom panel of Figure 1 shows the mixture of positive and negative
RM, reflecting the randomness of magnetic fields in the LSS.

With the coherence length of magnetic fields for RM (see Discussion)
expected to be smaller than the path length which should be a cosmological
scale, the inducement of RM is expected be a random walk
process.
Figure 2 shows the distributions of RM as well as other quantities along
a few LOS's through filaments;
it confirms that the inducement of RM is indeed a random walk process.
However, we note that the resulting RM is dominated by the contribution
from the density peak along LOS's.

To quantify RM in the LSS of the universe, we calculated the probability
distribution function (PDF) of $|{\rm RM}|$ for $512^2\times 3\times 16$
(${\rm projected\ grid\ zones} \times {\rm directions} \times {\rm runs}$)
LOS's.
Figure 3 shows the resulting PDF through the LOS's of different ranges
of the mean temperature weighted with X-ray emissivity, $T_X$.
The figure also shows the fitting to the log-normal distribution,
\begin{equation}
{\rm PDF}(\log_{10} |{\rm RM}|)=\frac{1}{\sqrt{2\pi\sigma^2}}
\exp\left[-\frac{(\log_{10} |{\rm RM}| - \mu)^2}{2\sigma^2}\right],
\end{equation}
finding that the PDF closely follows the log-normal distribution.
We also calculated the root mean square (rms) of RM,
$\langle{\rm RM}\rangle_{\rm rms}$;
note that the mean of RM, $\langle{\rm RM}\rangle$, is zero for our IGMF.
Through the WHIM, which mostly composes filaments,
$\langle{\rm RM}\rangle _{\rm rms}=1.41\ {\rm rad\ m^{-2}}$.
This agrees well with the value predicted with the mean strength and
coherence length of the IGMF in filaments by \citet{cr09}.
However, this is an order of magnitude smaller than the values of
$|{\rm RM}|$ toward the Hercules and Perseus-Pisces superclusters
reported in \citet{xkhd06}.
The difference is mostly due to the mass-weighted path length;
the value quoted by \citet{xkhd06} is about two orders of magnitude
larger than ours.
Through the hot gas with $10^7 < T < 10^8\ {\rm K}$,
$\langle{\rm RM}\rangle _{\rm rms}=108$ ${\rm rad\ m^{-2}}$, which is 
in good agreement with RM observations of galaxy clusters
\citep{cla01,cla04}.
Through the hot gas, however, we found RM of up to
$\ga 1000$ ${\rm rad\ m^{-2}}$.
This should be an artifact of limited resolution (see Discussion).
So the values for the hot gas in our work should not be taken seriously.

Finally, we calculated the two-dimensional power spectrum of RM on
$3\times 16$ (${\rm directions} \times {\rm runs}$) projected planes;
$P_{\rm RM}(k) \sim |{\rm RM}(\vec{k})|^2 k$, where ${\rm RM}(\vec{k})$
is the Fourier transform of ${\rm RM}(\vec{x})$ on planes.
Figure 4 shows the resulting power spectrum along with the power
spectra of electron density, magnetic fields, and the curl component of
flow motions, ${\vec v}_{\rm curl}$, which satisfies the relation
${\vec\nabla}\times{\vec v}_{\rm curl} \equiv {\vec\nabla}\times{\vec v}$.
The power spectrum of RM peaks at $k\sim 100$, which corresponds to
$\sim 1\ h^{-1}$ Mpc.
Cosmic variance is not significant around the peak, although it is
larger at smaller $k$, as expected.
The power spectrum of RM reflects the spatial distributions of
electron density, $n_e$, and LOS magnetic field, $B_{\parallel}$.
The power spectra of projected $n_e$ and projected $B_{\parallel}$,
have peaks at $\sim 3\ h^{-1}$ Mpc and
$\sim 1.5\ h^{-1}$ Mpc, respectively.
The shape of the power spectrum of RM follows that of projected
$B_{\parallel}$ rather than that of projected $n_e$, implying that
the statistics of RM would primarily carry the statistics of the IGMF.

\section{Discussion}

Our results depend of RM on the strength and coherence length of the IGMF.
We employed a model where the strength of the local IGMF was
estimated based on turbulence dynamo, while the direction was gripped
from structure formation simulations with passive fields (see Section 2).
In principle, if we had performed full MHD simulations, we could have followed 
the amplification of the IGMF through turbulence dynamo along with its
direction.
In practice, however, the currently available computational resources do not 
allow a numerical resolution high enough to reproduce the full development
of MHD turbulence.
Since the numerical resistivity is larger than the physical resistivity
by many orders of magnitude, the growth of magnetic fields is expected to
be saturated before dynamo action becomes fully operative
\citep[see, e.g.,][]{kcor97}.
In such situation, the state of magnetic fields in full MHD, including,
for instance, the power spectrum, is expected to mimic that of passive fields.
This is the reason why we adopted the model of \citet{rkcd08} to estimate
the strength of the IGMF, but we still used passive fields from structure
formation simulations to model the field direction.

The validity of our model IGMF was checked as follows:\hfill\break
1) In MHD turbulence, the distribution of magnetic fields, including the
direction, is expected to correlate with that of vorticity, since magnetic
fields and vorticity are described by similar equations except the
baroclinity term in the equation for vorticity (if dissipative processes
are ignored) \citep[see, e.g.,][]{kcor97}.
Such a correlation can be clearly seen in Figure 5, in which we depicts the 
distributions of our IGMF and vorticity in two-dimensional slices.\hfill\break
2) Full MHD turbulence simulations suggest that the peak of magnetic
field spectrum occurs $\sim 1/2$ of the energy injection scale, or the peak
scale of velocity power spectrum, at saturation; in the linear growth stage,
the peak scale of magnetic field spectrum grows as $\sim t^{1.5}$ or so
\citep{cr09}.
With our model IGMF, the peak scale of magnetic field spectrum is
$\sim 1\ h^{-1}$ Mpc (the third panel of Figure 4);
on the other hand, the curl component of flow motions has the peak of
power spectrum at $\sim 4\ h^{-1}$ Mpc (the bottom panel of Figure 4).
That is, the peak scale of magnetic field spectrum is $\sim 1/4$ of
the energy injection scale in our model IGMF.
By considering the turbulence in the LSS of the universe has not yet reached
the fully saturated stage \citep[see, e.g.,][]{rkcd08}, the ratio of the two
scales seems to be feasible.\hfill\break
These suggest that our model IGMF would produce reasonable results, although
eventually it needs to be replaced with that from full MHD simulations for
cosmological structure formation when computational resources allow such
simulations in future.

Apart from our model for the IGMF, the finite numerical resolution of
simulations could affect our results.
The average strength of our model IGMF is $\langle B\rangle \sim$ a few
$\mu$G in clusters/groups, $\sim 0.1 \mu$G around clusters/groups, and
$\sim 10$ nG in filaments.
\citet{rkcd08} tested the numerical convergence of the estimation.
With simulations of different numerical resolutions for cosmological
structure formation, it was shown that $\langle B\rangle$ of our model
IGMF for the WHIM with $10^5 < T <10^7\ {\rm K}$ would
approach the convergence value within a factor $\sim 2 - 3$ at the
resolution of $512^3$ grids (see Figure S5 of SOM of \citet{rkcd08}).

It is, on the other hand, rather tricky to assess the effect of finite
resolution on the coherence length of our model IGMF, because the
definition of coherence length for RM is not completely clear and
the estimation of coherence length, for instance, for the filament
IGMF alone is not trivial.
We tried to quantify coherence length in the following three ways:
1) We directly calculated the coherence length of $B_{\parallel}$,
that is, the length with the same sign of $B_{\parallel}$, along LOS's.
Figure 6 shows the PDF of the resulting coherence length through the
WHIM, which composes mostly filaments.
It peaks at the length of 3 zones corresponding to $586\ h^{-1}$ kpc.
2) We calculated 3/4 times the integral scale,
\begin{equation}
\frac{3}{4} \times2 \pi
\frac{\int P_B^{3D}(k)/k\ dk }{\int P_B^{3D}(k)\ dk},
\end{equation}
which is the coherence length defined for RM in the incompressible limit
(see Introduction), for the IGMF inside the whole computational box of
$(100\ h^{-1}{\rm Mpc})^3$ volume.
Here, $P_B^{3D}(k)$ is the three-dimensional power spectrum of magnetic
fields (the third panel of Figure 4).
We found the value to be $\sim 800\ h^{-1}$ kpc for our model IGMF.
3) We also calculated the largest energy containing scale in the
whole computational box, which is the peak scale of $kP_B^{3D}(k)$
(not shown).
It is $\sim 900\ h^{-1}$ kpc for our model IGMF.
Note that the latter two values include contributions from the IGMF in
filaments as well as in clusters, sheets, and voids.
All the three scales are comparable.
These length scales are $\sim$ 3 to 5 times larger than the
grid resolution of our simulations, $195\ h^{-1}$ kpc.

\citet{cr09} studied characteristic lengths in incompressible
simulations of MHD turbulence (see Introduction);
based on it, they predicted that the coherence length for RM
would be a few $\times\ 100\ h^{-1}$ kpc in filaments, while a few
$\times\ 10\ h^{-1}$ kpc in clusters.
With our grid resolution of $195\ h^{-1}$ kpc, the coherence length of
the IGMF in clusters should not be resolved and so our estimation of RM
for clusters should be resolution-affected, as pointed in Section 3.
On the other hand, while the predicted coherence length for the IGMF
in filaments is still larger than the grid resolution, the estimated
coherence length of $B_{\parallel}$ for the WHIM is a couple of times
larger than the prediction for filaments.
It could be partly due to the limited resolution in our simulations. 
However, as noted in Section 3, RM is dominantly contributed by the
density peak along LOS's (Figure 2).

The above statements indicate that our estimate of the RM through
filaments is expected to have uncertainties, especially due to limited
resolution of our simulations;
the error in our estimation could be up to a factor of several.

\section{Summary and Conclusion}

We studied RM in the LSS of the universe, focusing on RM through filaments;
simulations for cosmological structure formation were used and the model
IGMF of \citet{rkcd08} and \citet{cr09} based on turbulence dynamo was
employed.
Our findings are summarized as follows.
1) With our model IGMF, the rms of RM through filaments at the present
universe is $\sim 1\ {\rm rad\ m^{-2}}$.
2) The PDF of $|{\rm RM}|$ through filaments follows the log-normal
distribution.
3) The power spectrum of RM due to the IGMF in the local universe
peaks at a scale of $\sim 1\ h^{-1}$ Mpc.
4) Within the frame of our mode IGMF, we expect that the uncertainty
in our estimation for the rms of RM through filaments, due to the finite
numerical resolution of simulations, would be a factor of a few.

We note that our model does not include other possible contributions
to the IGMF, for instance, that from galactic black holes (AGN
feedbacks) \citep[see, e.g.,][]{kdlc01}.
So our model may be regarded as a minimal model, providing a baseline
for the IGMF.
With such contributions, the real IGMF might be somewhat stronger,
resulting in somewhat larger RM.

It has been suggested that future radio observatories such as LOFAR,
ASKAP, MeerKAT and SKA could detect the extragalactic RM of
$\sim 1\ {\rm rad\ m^{-2}}$ we predict \citep[see, e.g.,][]{beck09}.
However, it is known that the typical galactic foreground of RM is a
few tens and of order ten ${\rm rad\ m^{-2}}$ in the low and high
galactic latitudes, respectively \citep[see, e.g.,][]{sk80}.
So the detection of the extragalactic RM of $\sim 1\ {\rm rad\ m^{-2}}$
or so could be possible only after the galactic foreground is removed.
We note, however, that with filaments at cosmological distance, the
peak of the power spectrum of the RM due to the IGMF in filaments
would occur at small angular scales;
for instance, for a filament at a distance of $100\ h^{-1}$ Mpc,
the peak would occur at $\sim 0.5$ degree or so.
This is much smaller than the expected angular scale of the peak of
the galactic foreground, which would be around tens degree
\citep[see, e.g.,][]{fsss01}.
Then, it would be plausible to extract the signature of the RM of
$\sim 1\ {\rm rad\ m^{-2}}$ due to the IGMF in filaments.
We leave this issue and connecting our theoretical prediction of
RM in the LSS of the universe to observation for future studies.

\acknowledgments

TA was supported in part by National Research Foundation of Korea
(R01-2007-000-20196-0).
DR was supported in part by National Research Foundation of Korea
(K20901001400-09B1300-03210)

\clearpage

\begin{figure}
\epsscale{.50}
\plotone{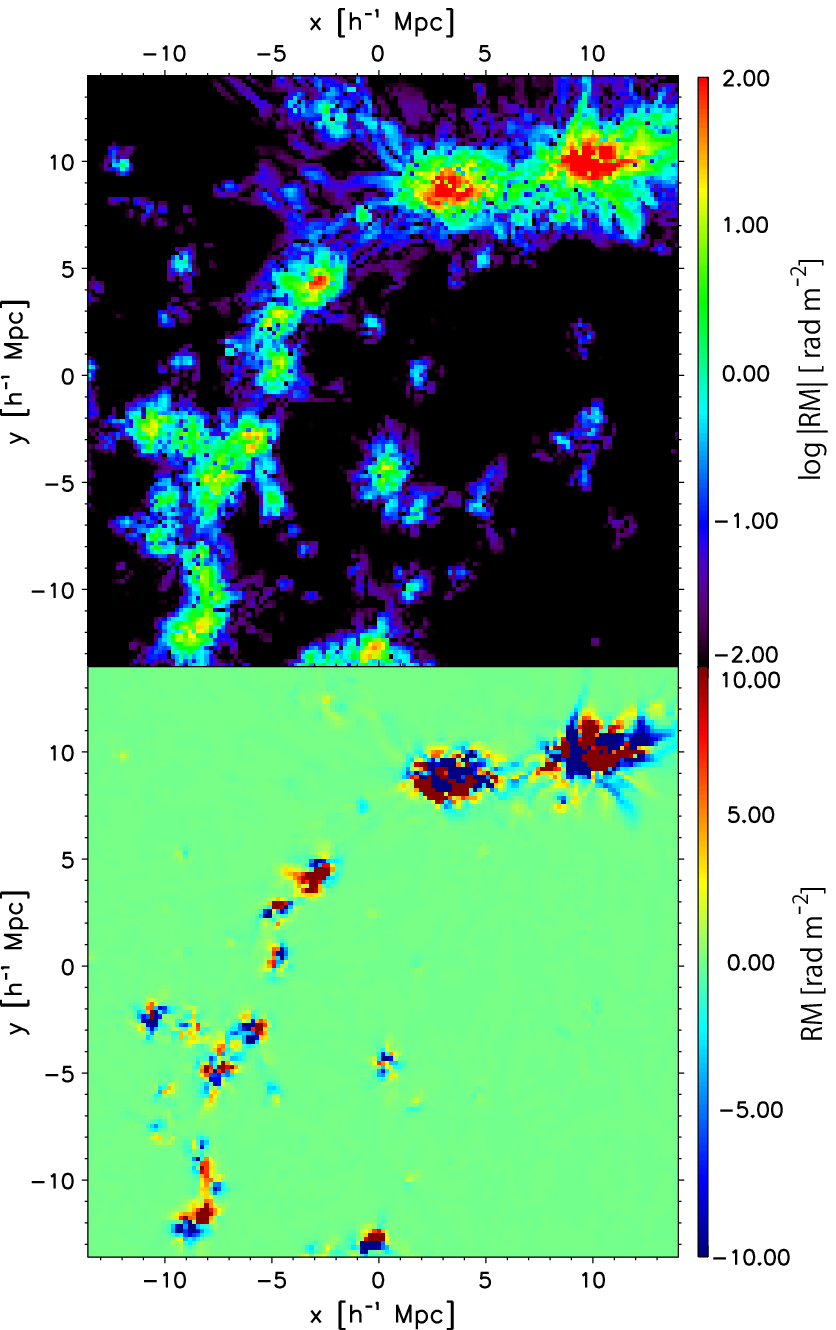}
\caption{RM map of $(28\ h^{-1}{\rm Mpc})^2$ area in the local universe
of depth (path length) of $L=100\ h^{-1}$.
Top and bottom panels show the map in logarithmic and linear scales,
respectively.}
\end{figure}

\clearpage

\begin{figure}
\epsscale{.90}
\plotone{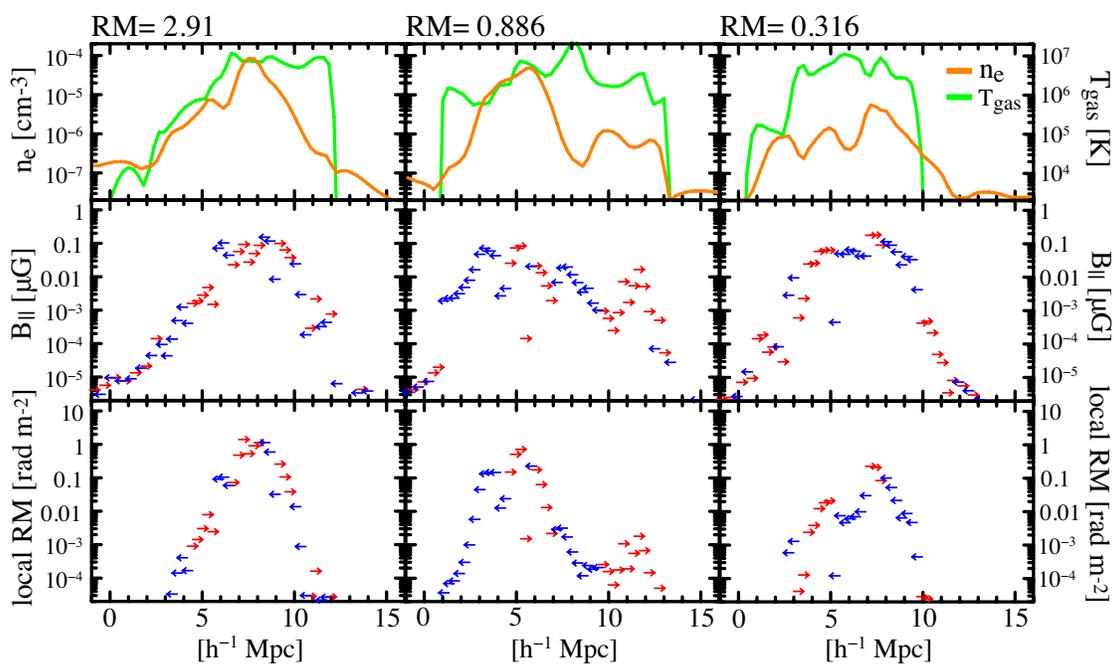}
\caption{Profiles of local RM, LOS magnetic field, $B_{\parallel}$,
electron density, $n_e$, and gas temperature, $T$, along a few LOS's
through filaments.
The arrows indicate the sign of local RM and $B_{\parallel}$.}
\end{figure}

\clearpage

\begin{figure}
\epsscale{.50}
\plotone{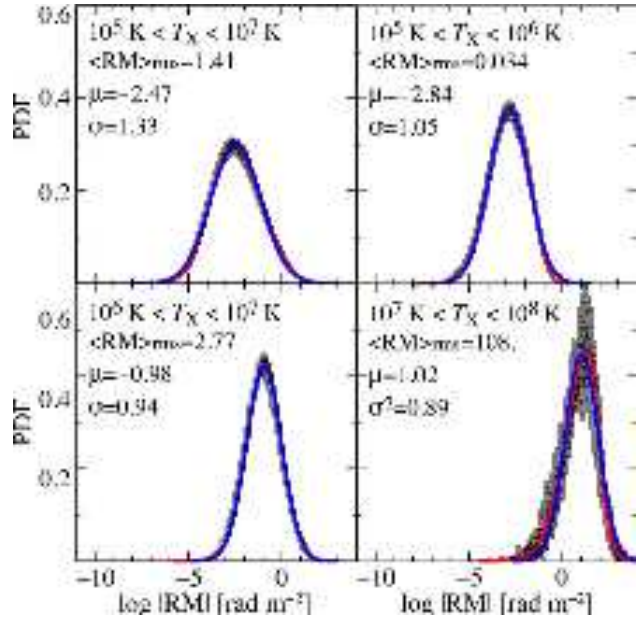}
\caption{PDF of $|{\rm RM}|$ through LOS's of different ranges
of the mean temperature weighted with X-ray emissivity, $T_X$.
Thin lines, thick lines (red or black), and thick lines (blue
or gray) show the PDFs from 16 independent runs, their average,
and the best-fit to the log-normal distribution, respectively.
The values of fitting parameters and the rms of RM are also shown.}
\end{figure}

\clearpage

\begin{figure}
\vskip -0.3cm
\epsscale{.4}
\plotone{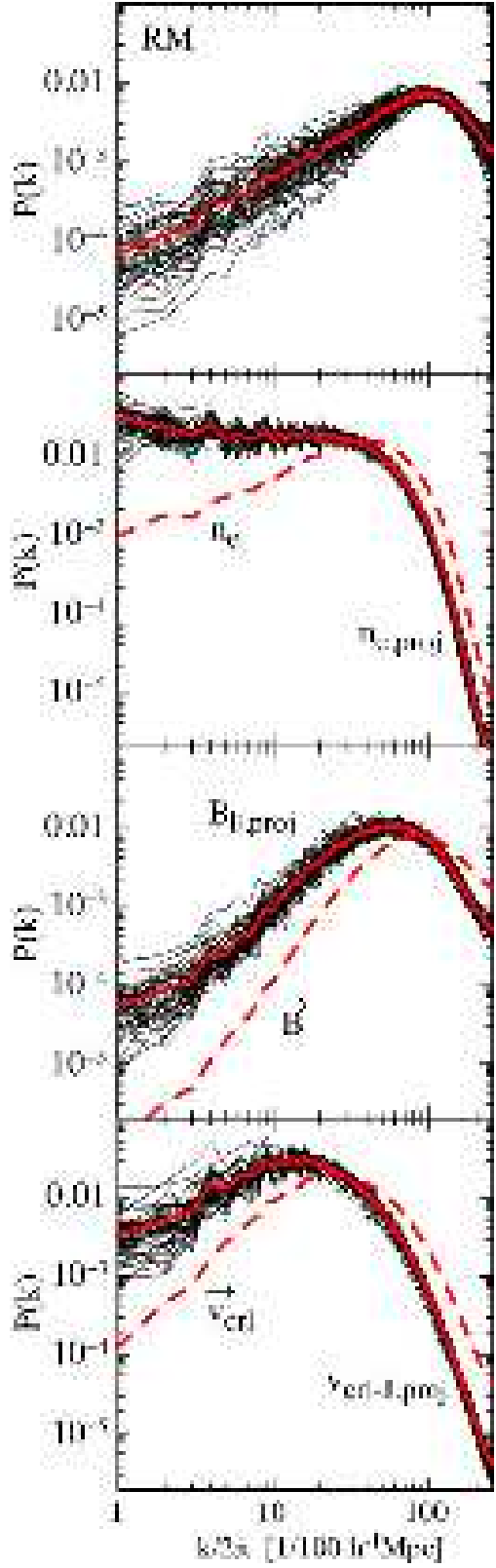}
\vskip -0.3cm
\caption{Two-dimensional power spectra of RM, projected electron density
($n_{\rm e,proj}$), projected line-of-sight IGMF strength
($B_{\parallel,{\rm proj}}$), and projected curl component of flow motions
(${\vec v}_{\rm curl}$) from top to bottom panels, respectively. 
Thin solid lines show the power spectra for $3\times 16$ two-dimensional
maps, and thick solid lines show their average.
The three-dimensional power spectra of $n_{\rm e}$, $\vec{B}$, and
${\vec v}_{\rm curl}$, are also shown with thick dashed lines.}
\end{figure}

\clearpage

\begin{figure}
\epsscale{1.0}
\plotone{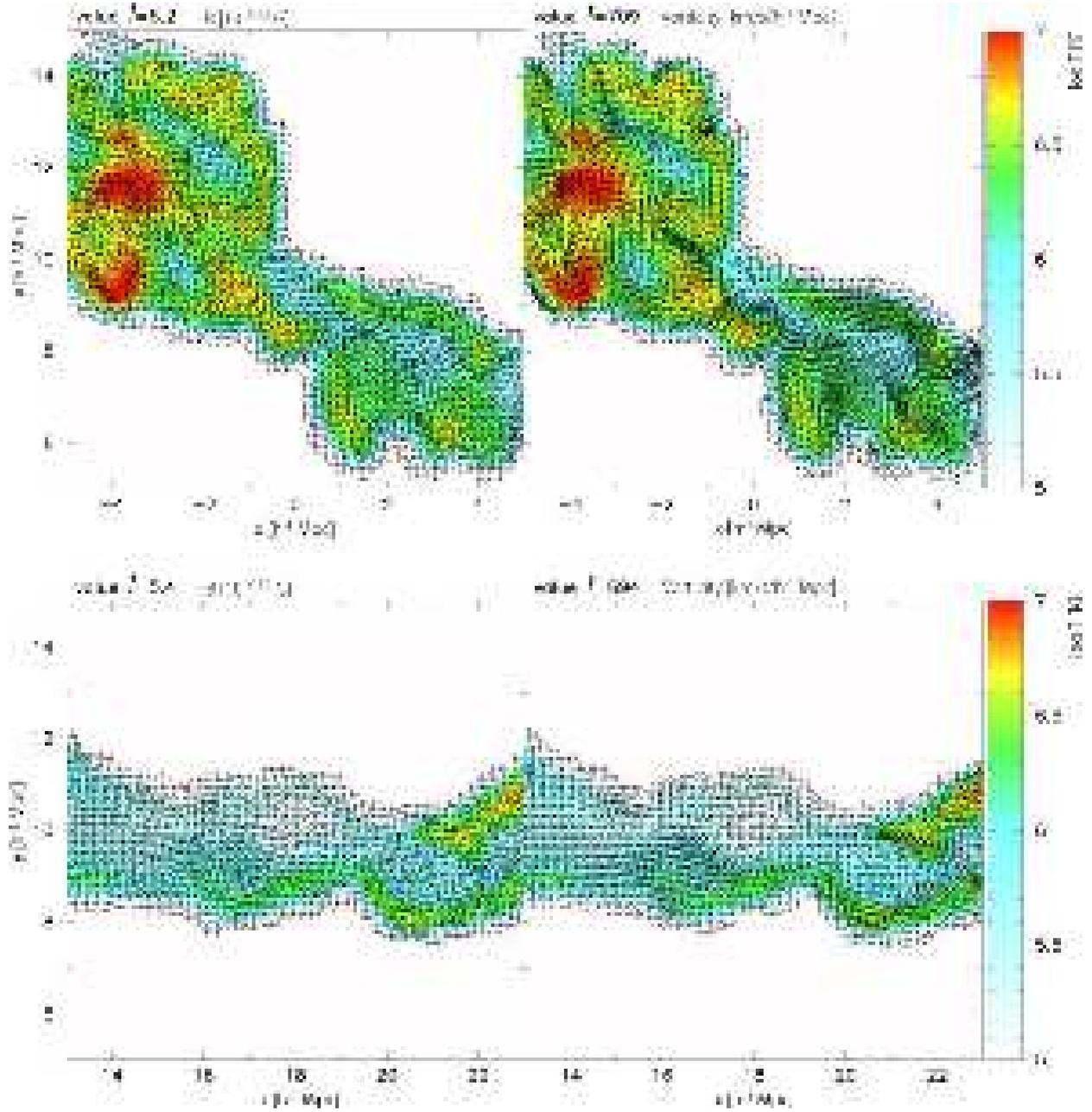}
\caption{Distributions of the IGMF (left) and vorticity (right) 
in two-dimensional slices. 
The length of arrows for the IGMF corresponds to $x$ of $10^{x-12}$~G. 
The color shows the gas temperature.
}
\end{figure}

\clearpage

\begin{figure}
\epsscale{.50}
\plotone{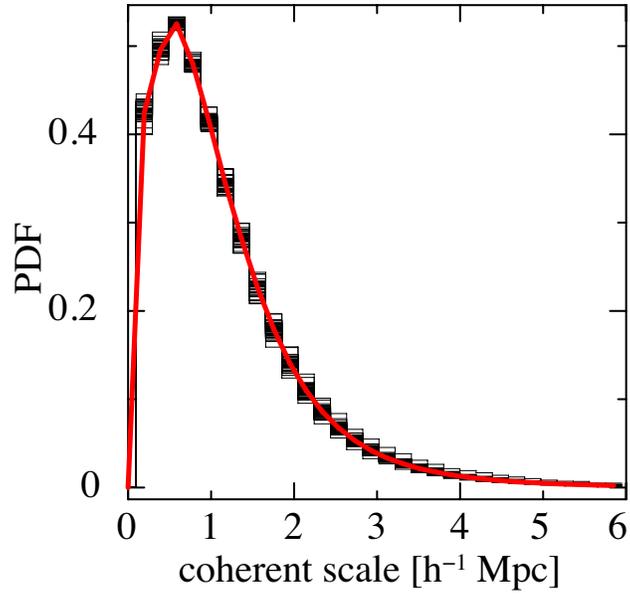}
\caption{PDF of the coherence length of $B_{\parallel}$ along LOS's
(the length with the same sign of $B_{\parallel}$) through the WHIM.
Thin and thick lines show the PDFs for 16 independent runs and their
average, respectively.}
\end{figure}

\end{document}